# Perpetual Assurances for Self-Adaptive Systems

Danny Weyns[1], Nelly Bencomo[2], Radu Calinescu[3], Javier Camara[4], Carlo Ghezzi[5], Vincenzo Grassi[6], Lars Grunske[7], Paola Inverardi[8], Jean-Marc Jezequel[9], Sam Malek[10], Raffaela Mirandola[5], Marco Mori[11] and Giordano Tamburrelli [12]

[1] Katholieke Universiteit Leuven Belgium, Linnaeus University Sweden, [2] Aston University UK, [3] University of York UK, [4] Carnegie Mellon University USA, [5] Politecnico di Milano Italy, [6] University of Rome Italy, [7] Humboldt University of Berlin Germany, [8] University of L'Aquila Italy, [9] IRISA France, [10] University of Irvine USA, [11] University of Florence Italy, [12] Vrije Universiteit Amsterdam The Netherlands
Contact: danny.weyns@cs.kuleuven.be

**Abstract.** Providing assurances for self-adaptive systems is challenging. A primary underlying problem is uncertainty that may stem from a variety of different sources, ranging from incomplete knowledge to sensor noise and uncertain behavior of humans in the loop. Providing assurances that the self-adaptive system complies with its requirements calls for an enduring process spanning the whole lifetime of the system. In this process, humans and the system jointly derive and integrate new evidence and arguments, which we coined *perpetual assurances* for self-adaptive systems. In this paper, we provide a background framework and the foundation for perpetual assurances for self-adaptive systems. We elaborate on the concrete challenges of offering perpetual assurances, requirements for solutions, realization techniques and mechanisms to make solutions suitable. We also present benchmark criteria to compare solutions. We then present a concrete exemplar that researchers can use to assess and compare approaches for perpetual assurances for self-adaptation.

## 1 Introduction

In recent years, researchers have started studying the impact of self-adaptation on the software engineering process [1][7][44]. This led to the insight that today's iterative, incremental, and evolutionary software engineering processes do not meet the requirements of many contemporary systems that need to handle change during system operation. In self-adaptive systems change activities are shifted from development time to runtime, and the responsibility for these activities is shifted from software engineers or system administrators to the system itself. Therefore the traditional boundary between development time and runtime blurs, which requires a complete reconceptualization of the software engineering process. An important aspect of the software engineering process of self-adaptive systems – in particular business or safety critical systems – is providing new evidence that the system requirements are satisfied during its entire lifetime, from inception to and throughout operation. This evidence must be produced despite the uncertainty that affects the requirements and/or the behavior of the environment, which lead to changes that may affect the system, its goals, and its environment [22][27]. To provide guarantees that the system requirements are satisfied, the state of the art in self-adaptive systems advocates the use of formal models as one promising approach, see e.g., [13][32][41][65][66][72]. Some approaches employ formal methods to provide guarantees by construction. More recently, the use of probabilistic models to handle uncertainties has gained interest. These models are used to verify properties and support decision-making about adaptation at runtime. However, providing assurances that the goals of self-adaptive systems are achieved during the entire life cycle remains a difficult challenge.

In this paper, we provide a background framework and the foundation for an approach to providing assurances for self-adaptive systems that we coined "perpetual assurances for self-adaptive systems." We elaborate on the challenges of perpetual assurances, requirements for solutions, realization techniques and mechanisms to make solutions suitable, and benchmark criteria to compare solutions. We then present a case study that researchers can use to assess, challenge and compare their approaches to providing assurances for self-adaptation. The paper concludes with a summary of the main challenges we identified to realize perpetual assurances for self-adaptive systems.

## 2  Key Challenges for Perpetual Assurances

We use the following working definition for *assurances for self-adaptive systems*:

> *Assurances for self-adaptive systems mean providing evidence for requirements compliance; this evidence can be provided off-line (i.e., not directly connected or controlled by the running system) and complemented online (i.e., connected or under control of the running system).*

We use the following working definition for *perpetual assurances for self-adaptive systems*:

> *Perpetual assurances for self-adaptive systems mean providing evidence for requirements compliance through an enduring process that continuously provides new evidence by combining system-driven and human-driven activities to deal with the uncertainties that the system faces across its lifetime, from inception to and throughout operation in the real world.*

Thus, providing assurances cannot be achieved by simply using off-line solutions possibly complemented with online solutions. Instead, we envisage that perpetual assurances will employ a continuous process where humans and the system jointly and continuously derive and integrate new evidence and arguments required to assure stakeholders (e.g., end users and system administrators) that the requirements are met by the self-adaptive system despite the uncertainties it faces throughout its lifetime.

Realizing the vision of perpetual assurances for self-adaptive systems poses multiple challenges, which we discuss next. From these challenges, we identify requirements, we overview relevant approaches for assurances and discuss mechanisms to make these approaches suitable for perpetual assurances, and finally we define benchmark criteria to compare different solutions.

A primary underlying challenge stems from *uncertainty*. The literature provides several definitions of uncertainty, ranging from *absence of knowledge* to *inadequacy of information* and *the difference between the amount of information required to perform a task and the amount of information already possessed* [5][35][51][54][62][67]. [62] and later [53] refer to uncertainty as *any deviation from the unachievable ideal of completely deterministic knowledge of the relevant system.* Such deviations can lead to a lack of confidence in the obtained results, based on a judgment that they might be incomplete, blurred, inaccurate, unreliable, inconclusive, or potentially false [55].

Hereafter, we make use of the taxonomy of uncertainty initially proposed in [53], where uncertainties are classified along three dimensions: *location*, *level,* and *nature*. The *location* dimension refers to where uncertainty manifests in the description (the model) of the studied system or phenomenon. We can specialize it into: (i) input parameters, (ii) model structure, and (iii) context. The *level* dimension rates how uncertainty occurs along a spectrum ranging from deterministic knowledge (level 0) to total ignorance (level 3). Awareness of uncertainty (level 1) and unawareness (level 2) are the intermediate levels. The *nature* dimension indicates whether the uncertainty is due to the lack of accurate information (*epistemic*) or to the inherent variability of the phenomena being described (*aleatory*). Other taxonomies of uncertainties related to different modeling aspects can be found [8][38][54].

Recognizing the presence of uncertainty and managing it can mitigate its potentially negative effects and increase the level of assurance in a given self-adaptive system. By ignoring uncertainties, one could draw unsupported claims on the system validity or generalize them beyond their bounded scope.

**Table 1 Classification of sources of uncertainty based on [50]**

| Source of Uncertainty | | Classification | |
|---|---|---|---|
| | | **Location** | **Nature** |
| Simplifying assumptions | System | Structural/context | Epistemic |
| Model drift | | Structural | Epistemic |
| Incompleteness | | Structural | Epistemic/Aleatory |
| Future parameters value | | Input | Epistemic |
| Adaptation functions | | Structural | Epistemic/Aleatory |
| Automatic learning | | Structural/input | Epistemic/Aleatory |
| Decentralization | | Context/structural | Epistemic |
| Requirements elicitation | Goals | Structural/input | Epistemic/Aleatory |
| Specification of goals | | Structural/input | Epistemic/Aleatory |
| Future goal changes | | Structural/input | Epistemic/Aleatory |
| Execution context | Context | Context/structural/ input | Epistemic |
| Noise in sensing | | Input | Epistemic/Aleatory |
| Different sources of information | | Input | Epistemic/Aleatory |
| Human in the loop | Human | Context | Epistemic/Aleatory |
| Multiple ownership | | Context | Epistemic/Aleatory |

**Table 1** shows a classification of uncertainty based on [50], which is inspired by the literature on self-adaptation (e.g., [30][36].) Each source of uncertainty is classified according to the *location* and *nature* dimensions of the taxonomy. The *level* dimension of the taxonomy is not shown in the table since each source of uncertainty can be of any level depending on the implemented capabilities in the system that should deal with the uncertainty.

Sources of uncertainty are structured in four groups: (i) sources of uncertainty related to the *system* itself; (ii) uncertainty related to the system *goals*; (iii) uncertainty in the execution *context*; and (iv) uncertainty related to *human* aspects. We provide a brief definition of each type of uncertainty source, focusing on *incompleteness, automated learning,* and *multiple ownership,* which especially apply to self-adaptive systems.

**Sources of uncertainty related to the system itself:**

- *Simplifying assumptions*: the model, being an abstraction, includes per-se a degree of uncertainty, since details whose significance is supposed to be minimal are not included in the model.
- *Model drift*: as the system adapts, its structure also changes, so it is necessary to keep aligned systems and corresponding models to avoid reasoning on outdated models.
- *Incompleteness* manifests when some parts of the system or its model are knowingly missing. Incompleteness can show up at design time and also at runtime. Progressive completion at design time occurs in iterative, exploratory development processes. Completion at runtime occurs, for example, in dynamic service discovery and binding in service-oriented systems. Managing incompleteness is challenging. Performing assurance on incomplete systems requires an ability to discover which sub-assurances must be delayed until completion occurs, and this may even happen at runtime [23]. Consequently, assurance must be incremental and comply with the time constraints imposed by the runtime environment.
- *Future parameter value*: uncertainty in the future world where the system will execute creates uncertainties in the correct actions to take at present.
- *Automatic learning* in self-adaptive systems usually uses statistical processes to create knowledge about the execution context and most-useful behaviors. These learning processes can lead to applications with uncertain behavior. The location of this uncertainty may be in the model structure or input parameters, depending on how general the concepts are for which the application has been provided with machine-leaning capabilities. The nature of the uncertainty depends on the point of view. From the viewpoint of the application and its models of the world, as the machine has to learn from imperfect and limited data, the nature of the uncertainty is epistemic. From the user viewpoint, since the information undergoes a statistical learning process during model synthesis, automatic learning introduces randomness in the model and analysis results, and consequently it can be seen as an aleatory uncertainty.
- *Decentralization*: this uncertainty comes from the difficulty to keep a model updated if the system is decentralized and each subsystem has self-adaptation capabilities. The reason is that it can be hardly expected that any subsystem could obtain accurate knowledge of the state of the entire system.

**Sources of uncertainty related to the goals of the system:**

- *Elicitation of requirements*: identifying the needs of stakeholders is known to be problematic in practice. Issues concerns defining the system scope, problems in fostering understanding among the different communities with a stake in the system, and problems in dealing with the volatile nature of requirements. The latter issue is particularly relevant for self-adaptation.
- *Specification of goals*: accurately specifying the preferences of stakeholders is difficult and prone to uncertainty. For example, the runtime impact of a utility function that defines the desired tradeoffs between conflicting qualities may not be completely known upfront.
- *Future goal changes*: the requirements of the system can change due to new needs of the customers, to new regulations or to new market rules, etc.

**Sources of uncertainty related to the execution context of the system:**

- *Execution context*: the context in which the application operates should be represented in the model, but the available monitoring mechanisms might not suffice to univocally determine this context and its evolution over time.
- *Noise in sensing*: the sensors/monitors are not ideal devices and they can provide, for example, slightly different data in successive measures while the actual value of the monitored data did not change.

**Sources of uncertainty related to the humans aspects:**

- *Human in the loop*: human behavior is intrinsically uncertain and systems commonly rely on correct human operations – both as system users and as system administrators. This assumption of correct operations may not hold because human behavior can diverge from the correct behavior.
- *Multiple ownership*: systems are increasingly assembled out of parts provided by different stakeholders, whose exact nature and behavior may be partly unknown when the system is composed. This is typical of service-oriented, dynamically composed systems, and systems of systems. The nature of this uncertainty is mainly epistemic. However, it may also be aleatory since third-party components may change without notifying the owner of the integrated application.

As discussed above, uncertainty in its various forms represents as the ultimate source of both motivations for and challenges to perpetual assurance. Uncertainty manifests through *changes*. For example, uncertainty in capturing the precise behavior of an input phenomenon to be controlled results in assumptions made to implement the system. Therefore, the system must be calibrated later when observations of the physical phenomenon are made. This in turn leads to changes in the implemented control system that must be scrutinized for assurances.

## 3 Requirements for Solutions to Realize Perpetual Assurances

The provision of perpetual assurance for self-adaptive systems must satisfy additional requirements compared to those met by assurance solutions for non-adaptive systems. These additional requirements correspond to the challenges summarized in the previous section. In particular, perpetual assurance solutions must cope with a variety of uncertainty sources that depend on the purpose of self-adaptation and the environment in which the assured self-adaptive system operates. To this end, they must build and continually update their assurance arguments through integrating two types of assurance evidence. The first type of evidence corresponds to system and environment elements not affected significantly by uncertainty, and therefore can be obtained using traditional offline approaches [47]. The second type of evidence is associated with the system and environment components affected by the sources of uncertainty summarized in **Table 1**. These requirements basically refer to the different functions of adaptation: from sensing to monitoring, analyzing, planning, executing, and activating [23]. The evidence must be synthesized at runtime, when the uncertainty is treated, i.e., be reduced, quantified, or resolved sufficiently, to enable such synthesis. The requirements described below stem from this need to treat uncertainty, to generate new assurance evidence, and to use it to update existing assurance arguments.

**Requirement 1**: *A perpetual assurance solution must continually observe the sources of uncertainty affecting the self-adaptive system*. Without collecting data about the relevant sources of uncertainty, the solution cannot acquire the knowledge necessary to provide the assurance evidence unachievable offline.[1] Addressing this requirement may necessitate, for instance, the instrumentation of system components, the sensing of environmental parameters, and the monitoring of user behavior.

**Requirement 2:** *A perpetual assurance solution must use its observations of uncertainty sources to continually quantify and potentially reduce the uncertainties affecting its ability to provide assurance evidence*. The raw data obtained through observing the sources of uncertainty need to be used to reduce the uncertainty (e.g., through identifying bounds for the possible values of an uncertain parameter). This uncertainty quantification and potential reduction must be sufficient to enable the generation of the assurance evidence that could not be obtained offline. For instance, rigorous learning techniques such as those introduced in [15][17][28][29] can be used to continually update formal models whose runtime analysis generates new assurance evidence.

**Requirement 3:** *A perpetual assurance solution must continually deal with overlapping sources of uncertainty, and may need to treat these sources in a compose-able fashion*. The sources of uncertainty from **Table 1** hardly ever occur apart or independently from each other. Instead, their observed effects are often compounded, consequently inhibiting the system from clearly assessing the extent to which it satisfies its requirements. Therefore, solutions, assurance arguments and the evidence they build upon need to be able to tackle composed sources of uncertainty.

**Requirement 4:** *A perpetual assurance solution must continually derive new assurance evidence*. Up-to-date knowledge that reduces the uncertainty inherent within the self-adaptive system and its environment must be used to infer the missing assurance evidence (e.g., to deal with model drift or handle goal changes as described in Section 2). As an example, new evidence can be produced through formal verification of updated models that reflect the current system behavior [11][12][13][14][32]. Complementary approaches to verification include online testing, simulation, human reasoning and combinations thereof.

**Requirement 5:** *A perpetual assurance solution must continually integrate new evidence into the assurance arguments for the safe behavior of the assured self-*

---

[1] If all necessary assurance evidence can be derived using knowledge available offline, then we

*managing system*. Guaranteeing safe behavior requires the provision of assurance arguments that, in the case of self-adaptive systems, must combine offline evidence with new evidence obtained online, once uncertainties are resolved to the extent that enables the generation of the new evidence.

**Requirement 6:** *A perpetual assurance solution may need to continually combine new assurance evidence synthesized automatically and provided by human experts.* Developing assurance arguments is a complex process that is carried out by human experts and includes the derivation of evidence through both manual and automated activities. Online evidence derivation will in general require a similar combination of activities, although it is conceivable that self-adaptive systems affected by reduced levels and sources of uncertainty might not always be able to assemble all the missing evidence automatically.

**Requirement 7:** *A perpetual assurance solution must provide assurance evidence for the system components, the human activities and the processes used to meet the previous set of requirements.* As described above, perpetual assurances for self-adaptive systems cannot be achieved without employing system components and a mix of automated and manual activities not present in the solutions for assuring non-adaptive systems. Each of these new elements will need to be assured either by human-driven or machine-driven approaches. As an example, a model checker used to obtain new assurance evidence arguments at runtime will need to be certified for this type of use.

In addition to the functional requirements summarized so far, the provision of perpetual assurance must be timely, non-intrusive and auditable, as described by the following non-functional requirements.

**Requirement 8:** *The activities carried out by a perpetual assurance solution must produce timely updates of the assurance arguments.* The provision of assurance arguments and of the evidence underpinning them is time consuming. Nevertheless, perpetual assurance solutions need to complete updating the assurance arguments guaranteeing the safety of a self-adaptive system timely, before the system engages in operations not covered by the previous version of the assurance arguments.

**Requirement 9:** *The activities of the perpetual assurance solution, and their overheads must not impact the operation of the assured self-adaptive system.* The techniques employed by a perpetual assurance solution (e.g., instrumentation, monitoring and online learning, and verification) must not interfere with the ability of the self-adaptive system to deliver its intended functionality effectively and efficiently.

**Requirement 10:** *The assurance evidence produced by a perpetual assurance solution and the associated assurance arguments must be auditable by human stakeholders.* Assurance arguments and the evidence they build upon are intended for human experts responsible for the decision to use a system, for auditing its safety, and for examining what went wrong after failures. In keeping with this need, perpetual assurance solutions must express their new assurance evidence in human-readable format.

**Table 2** summarizes the requirements for perpetual assurance solutions.

**Table 2 Summary requirements**

| Requirement | Brief description |
|---|---|
| R1: Monitor uncertainty | A perpetual assurance solution must continually observe the sources of uncertainty affecting the self-adaptive system. |
| R2: Quantify uncertainty | A perpetual assurance solution must use its observations of uncertainty sources to continually quantify and potentially reduce the uncertainties affecting its ability to provide assurance evidence. |
| R3: Manage overlapping uncertainty sources | A perpetual assurance solution must continually deal with overlapping sources of uncertainty and may need to treat these sources in a composable fashion. |
| R4: Derive new evidence | A perpetual assurance solution must continually derive new assurance evidence arguments. |
| R5: Integrate new evidence | A perpetual assurance solution must continually integrate new evidence into the assurance arguments for the safe behavior of the assured self-managing system. |
| R6: Combine new evidence | A perpetual assurance solution may need to continually combine new assurance evidence synthesized automatically and provided by human experts. |
| R7: Provide evidence for the elements and activities that realize R1-R6 | A perpetual assurance solution must provide assurance evidence for the system components, the human activities, and the processes used to meet the previous set of requirements. |
| R8: Produce timely updates | The activities carried out by a perpetual assurance solution must produce timely updates of the assurance arguments. |
| R9: Limited overhead | The activities of the perpetual assurance solution and their overheads must not impact the operation of the assured self-adaptive system. |
| R10: Auditable arguments | The assurance evidence produced by a solution and the associated assurance arguments must be auditable by human stakeholders. |

## 4 Approaches to Realize Perpetual Assurances

Several approaches have been developed in the previous decades to check whether a software system complies with its requirements. Important relevant research in this context is models at runtime [4][9]. In this section, we give a brief overview of representative approaches. In the next section, we elaborate on mechanisms to make these approaches suitable to support perpetual assurances for self-adaptive systems, driven by the requirements described in the previous section. **Table 3** gives an overview of the approaches we discuss, organized in three groups: human-driven approaches (manual), system-driven (automated), and hybrid (manual and automated).

**Table 3 Approaches for assurances**

| Group | Approaches |
|---|---|
| Human-driven approaches | - Formal proof<br>- Simulation |
| System-driven approaches | - Runtime verification<br>- Sanity checks<br>- Contracts |
| Hybrid approaches | - Model checking<br>- Testing |

### 4.1 Human-Driven Approaches

We discuss two representative human-driven approaches for assurances: formal proof, and simulation.

**Formal proof** is a mathematical calculation to prove a sequence of connected theorems or formal specifications of a system. Formal proofs are conducted by humans with the help of interactive or automatic theorem provers, such as Isabelle or Coq. Formal proofs are rigorous and unambiguous, but require from the user both detailed knowledge about how the self-adaptive system works and significant mathematical experience. Furthermore, formal proofs only work for highly abstract models of a self-adaptive system (or more detailed models of small parts of it). Still the approach can be useful to build up assurance evidence arguments about basic properties of the system or specific elements (e.g., algorithms). We give two examples of formal proof used in self-adaptive systems. In [68], the authors formally prove a set of theorems to assure atomicity of adaptations of business processes in cross-organizational collaborations. The approach is illustrated in an example where a supplier wants to evolve its business process by changing the policy of payments. In [71][72], the authors formally prove a set of theorems to assure safety and liveness properties of self-adaptive systems. The approach is illustrated for data stream elements that modify their behavior in response to external conditions through the insertion and removal of filters.

**Simulation** comprises three steps: the design of a model of a system, the execution of that model, and the analysis of the output. Different types of models at different levels of abstraction can be used, including declarative, functional, constraint, spatial or multi-model. A key aspect of the execution of a model is the way time is treated, e.g., small time increments or progress based on events. Simulation allows a user to explore many different states of the system, without being prevented by an infinite (or at least very large) state space. Simulation runs can provide assurance evidence arguments at different levels of fidelity, which are primarily based on the level of abstraction of the model used and the type of simulation applied. As an example, in [19] the authors simulate self-adaptive systems and analyze the gradual fulfillment of requirements. The approach provides valuable feedback to engineers to iteratively improve the design of self-adaptive systems.

### 4.2 System-Driven Approaches

We discuss three representative system-driven approaches for assurances: runtime verification, sanity checks, and contracts.

**Runtime verification** is a well-studied lightweight verification technique that is based on extracting information from a running system to detect whether certain properties are violated. Verification properties are typically expressed in trace predicate formalisms, such as finite state machines, regular expressions, and linear temporal logics. Runtime verification is less complex than traditional formal verification approaches, such as model checking, because only one or a few execution traces are analyzed. As an example, in [56] the authors introduce an approach to estimate the probability that a temporal property is satisfied by a run of a program. The approach models event sequences as observation sequences of a Hidden Markov Model (HMM) and uses an algorithm for HMM state estimation, which determines the probability of a state sequence.

**Sanity checks** are calculations that check whether certain invariants hold at specific steps in the execution of the system. Sanity checks have been used for decades because they are very simple to implement and can easily be disabled when needed to address performance concerns. On the other hand, sanity checks provide only limited assurance evidence arguments that the system is not behaving incorrectly. As an example, in [59], the authors present an approach to manage shared resources that is based on an adaptive arbitrator that uses device and workload models. The approach uses sanity checks to evaluate the correctness of the decisions made by a constraint solver.

**Contracts.** The notion of Contract (introduced by B. Meyer in Design by Contract in Eiffel) makes the link between simple sanity checks and more formal, precise and verifiable interface specifications for software components. Contracts extend the ordinary definition of abstract data types with pre-conditions, post-conditions and invariants. Contracts can be checked at runtime as sanity checks, but can also serve as a basis for assume-guarantee modular reasoning. In [10] the authors have shown that the notion of functional contract as defined in Eiffel can be articulated in four layers: syntactic, functional, synchronization and quality of service contracts. For each layer, both runtime checks and assume-guarantee reasoning are possible.

### 4.3 Hybrid Approaches

We discuss two representative hybrid approaches to provide assurances: model checking and testing.

**Model checking** is an approach allowing designers to check that a property (typically expressed in some logic) holds for all reachable states in a system. Model checking is typically run offline and can only work in practice on either a high level abstraction of an adaptive system or on one of its components, provided it is simple enough. For example, in [25], the authors model MAPE loops of a mobile learning application in timed automata and verify robustness requirements expressed in timed computation tree logic using the Uppaal tool. The models and properties are concrete instances of a set of reusable formal templates for specifying MAPE-K loops [26]. To deal with uncertainty at design time and the practical coverage limitations of model checking for realistic system, several researchers have studied the application of model checking techniques at runtime. E.g., in [13] QoS requirements of service-based systems are expressed as probabilistic temporal logic formulae, which are then automatically analyzed at runtime to identify and enforce optimal system configurations. The approach in [43] applies model checking techniques at runtime with the aim of validating compliance with requirements (expressed in LTL) of evolving code artifacts.

**Testing** allows designers to check that the system performs as expected for a finite number of situations. Inherently testing cannot guarantee that a self-adaptive system fully complies with its requirements. Testing is traditionally performed before deployment. As an example, in [18] the authors introduce a set of robustness tests, which provide mutated inputs to the interfaces between the controller and the target system, i.e. probes. The set of robustness tests are automatically generated by applying a set of predefined mutation rules to the messages sent by probes. Traditional testing allows demonstrating that a self-adaptive system is capable of satisfying particular requirements before deployment. However, it cannot deal with unanticipated system and environmental conditions. Recently, researchers in self-adaptive systems have started investigating how testing can be applied at runtime to deal with this challenge. In [34], the authors motivate the need and identify challenges for the testing of self-adaptive systems at runtime. The paper introduces the so-called MAPE-T feedback loop for supplementing testing strategies with runtime capabilities based on the monitoring, analysis, planning, and execution architecture for self-adaptive systems.

## 5 Mechanisms to Make Perpetual Assurances Working

Turning the approaches of perpetual assurances into a working reality requires tackling several issues. The key challenge we face is to align the approaches with the requirements we have discussed in Section 1.2. For the functional requirements (R1 to R7), the central problem is how to collect assurance evidence arguments at runtime and compose them with the evidence acquired throughout the lifetime of the system possibly by different approaches. For the quality requirements (R8 to R10) the central problem is to make the solutions efficient and support the interchange of evidence arguments between the system and users. In this section, we first elaborate on the implications of quality requirements on approaches for perpetual assurances. Then we discuss two classes of mechanisms that can be used to provide the required functionalities for perpetual assurances and meet the required qualities: decomposition mechanisms and model-driven mechanisms.

### 5.1 Quality Properties for Perpetual Assurances Approaches

The efficiency of the approaches for perpetual assurances must be evaluated with respect to the size of the self-adaptive system (e.g., number of components in the system) and the dynamism it is subject to (e.g., the frequency of change events that the system has to face). An approach for perpetual assurances of self-adaptive systems is efficient if, for systems of realistic size and dynamism, it is able to:

- Provide results (assurances) within defined time constraints (depending on the context of use);
- Consume a limited amount of resources, so that the overall amount of consumed resources over time (e.g. memory size, CPU, network bandwidth, energy, etc.) remains within a limited fraction of the overall resources used by the system.
- Scale well with respect to potential increases in size of the system and the dynamism it is exposed to.

In the next sections, we discuss some promising mechanisms to make approaches for perpetual assurances work. Before doing so, it is important to note that orthogonal to the requirements for perceptual assurance in general, the level of assurance that is needed depends on the requirements of the self-adaptive system under consideration. In some cases, combining regular testing with simple and time-effective runtime techniques, such as sanity checks and contract checking, will be sufficient. In other cases, more powerful approaches are required. For example, model checking could be used to verify a safe envelope of possible trajectories of an adaptive system at design time, and verification at runtime to check whether the next change of state of the system keeps it inside the pre-validated envelope.

### 5.2 Decomposition Mechanisms for Perpetual Assurances Approaches

The first class of promising mechanisms for the perpetual provisioning of assurances is based on the principle of decomposition, which can be carried out along two dimensions:

1. Time decomposition, in which:
   a. Some preliminary/intermediate work is performed off-line, and the actual assurance is provided on-line, building on these intermediate results;
   b. Assurances are provided with some degree of approximation/coarseness, and can be refined if necessary.
2. Space decomposition where verification overhead is reduced by independently verifying each individual component of a large system, and then deriving global system properties through verifying a composition of its component-level properties. Possible approaches that can be used to this end are:
   a. Flat approaches, that only exploit the system decomposition into components;
   b. Hierarchical approaches, where the hierarchical structure of the system is exploited;
   c. Incremental approaches targeted at frequently changing systems, in which re-verification are carried out only on the minimal subset of components affected by a change.

We provide examples of state of the art approaches for each type.

#### 5.2.1 Time Decomposition

We consider two types of time decomposition: i) partially offline, assurances provided online, and ii) assurances with some degree of approximation.

**Time decomposition: partially offline, assurances provided online.** In [32] and [20], the authors propose approaches based on the pre-computation at design time of a formal and parameterized model of the system and its requirements; the model parameters are then instantiated at runtime, based on the output produced by a monitoring activity. Parametric probabilistic model checking [40] enables the evaluation of temporal properties on Parametric Markov Chains, yielding polynomials or rational functions as evaluation results. Instantiating the parameters in the result function, results for quantitative properties, such as probabilities or rewards can be cheaply obtained at runtime, based on a more computationally costly analysis process carried out offline. Due to limitations in practice of the parametric engines implemented in tools such as PARAM [40] or PRISM [46], the applicability of this technique may provide better results in self-adaptive systems with static architectures (or with a limited number of possible configurations), since dynamic changes to the architecture might require the generation of parametric expressions for new configurations.

**Time decomposition: assurances with some degree of approximation.** Statistical model checking [39][49][69] is a kind of simulation approach that enables the evaluation of probability and reward-based quantitative properties, similarly to probabilistic model checking. Specifically, the approach involves simulating the system for finitely many executions and using statistical hypothesis testing to determine whether the samples constitute statistical evidence of the satisfaction of a probabilistic temporal logic specification, without requiring the construction of an explicit representation of the state space in memory. Moreover, this technique provides an interesting tradeoff between analysis time and the accuracy of the results obtained (e.g., by controlling the number of sample system executions generated) [70]. This makes it appropriate for systems with strict time constraints in which accuracy is not a primary concern. Moreover, dynamic changes to the system architecture do not penalize resource consumption (e.g., no model reconstruction is required), making statistical model checking a good candidate for the provision of assurances in highly dynamic systems. Initial research that uses statistical techniques at runtime for providing guarantees in self-adaptive systems is reported in [42][64].

### 5.2.2 Space Decomposition

We consider three types of space decomposition: i) flat approaches, ii) hierarchical approaches, and iii) incremental approaches.

**Space decomposition: flat approaches.** In [47], the authors use assume-guarantee model checking to verify component-based systems one component at a time, and employs regular safety properties both as the assumptions made about system components and the guarantees they provide to reason about probabilistic systems in a scalable manner. [52] employs probabilistic interface automata to enable the isolated analysis of probabilistic properties in environment models (based on shared system/environment actions), which are preserved in the composition of the environment and a non-probabilistic model of the system.

**Space decomposition: hierarchical approaches.** The authors of [45] and [48] consider the decomposition of models to carry out incremental quantitative verification. This strategy enables the reuse of results from previous verification runs to improve efficiency. While [48] considers decomposition of Markov Decision Process-based Models into their strongly connected components (SCCs), [45] employs high-level algebraic representations of component-based systems to identify and execute the minimal set of component-wise re-verification steps after a system change.

**Space decomposition: incremental approaches.** In [16], the authors use assume-guarantee compositional verification to efficiently re-verify safety properties after changes that match some particular patterns, such as component failures.

*5.2.3 Discussion*

The discussed decomposition mechanisms lend themselves to the identification of possible division of responsibilities between human-driven and system-driven activities, towards the ultimate goal of providing guarantees that the system goals are satisfied. As an example, considering time decomposition approaches based on parametric model checking, the adopted parametric models can result from a combination of human-driven activities and artifacts automatically produced from other sources or information (architectural models, adaptation strategy specifications, etc.). Similarly, in case of space decomposition approaches based on assume-guarantee compositional verification, the assumptions used in the verification process could be automatically "learned" by the system [58], or provided by (or combined with input from) human experts.

**5.3 Model-based Mechanisms for Perpetual Assurances Approaches**

For whatever division of responsibilities between human and systems in the perpetual assurance process, an important issue is how to define in a clean, well-documented and traceable way the interplay between the actors involved in the process. Model-driven mechanisms can be useful to this end, as they can support the rigorous development of a self-adaptive system from its high-level design up to its running implementation. Moreover, they can support the controlled and traceable modification by humans of parts of the system and/or of its self-adaptive logic, e.g. to respond to modifications of the requirements to be fulfilled. In this direction [61] presents a model-driven approach to the development of adaptation engines. This contribution includes the definition of a domain-specific language for the modeling of adaptation engines, and a corresponding runtime interpreter that drives the adaptation engine operations. Moreover, the approach supports the combination of on-line machine controlled adaptations and off-line long-term adaptations performed by humans to maintain and evolve the system.

Steps towards the definition of mechanisms that can support the human-system interplay in the perpetual assurance process can also be found in [41]. The authors propose an approach called ActivFORMS in which a formal model of the adaptation engine (MAPE-K feedback loop) based on timed automata and adaptation requirements expressed in timed computation tree logic are complemented by a suitable virtual machine that can execute the models, hence guaranteeing at runtime the compliance of properties verified offline. The approach allows dynamically checking the adaptation requirements, complementing the evidence arguments derived from human-driven offline activities with arguments provided online by the system that could not be obtained offline (e.g., for performance reasons). The approach supports deployment of new models at runtime elaborated by human experts to deal with new goals.

# 6 Benchmark Criteria for Perpetual Assurances

This section provides benchmark criteria to compare different approaches that provide perpetual assurances. We identify benchmark criteria along four aspects: capabilities of approaches to provide assurances, basis of evidence for assurances, stringency of assurances, and performance of approaches to provide assurances. The criteria cover both functional and quality requirements for perpetual assurances techniques.

## 6.1 Capabilities of Approaches to Provide Perpetual Assurances

The first benchmark aspect compares the extent to which approaches differ in their ability to provide assurances for the requirements (goals) of self-adaptation. We distinguish the following benchmark criteria: variability (in requirements and the system itself), inaccuracy & incompleteness, conflicting criteria, user interaction, and handling alternatives.

**Variability.** Support for variability of requirements is necessary in ever-changing environments where requirements are not statically fixed, but are runtime entities that should be accessible and modifiable. An effective assurance approach should take into account requirements variations, including addition, updating, and deletion of requirements. For instance, variations to service-level agreement (SLA) contract specifications may have to be supported by an assurance approach. Variations may also involve the system itself, in term of unmanaged modifications; that is, an assurance approach should support system variability in terms of adding, deleting, and updating managed system components and services. This process may be completely automatic or may involve humans.

**Inaccuracy & incompleteness.** The second benchmark criterion deals with the capability of an assurance approach to provide evidence with models that are not 100% precise and complete. Inaccuracy may refer to both system and context models. For instance, it may not be possible to accurately define requirements at design time, thus making it necessary to leave a level of uncertainty within the requirements. On the other hand context models representing variations to context elements may be inaccurate, which may affect the results of an assurance technique. Similarly, incompleteness may either concern system or context models. For instance, either new services or new context elements may appear only at runtime. An assurance technique should provide the best possible assurance level regarding inaccurate and incomplete models. Additionally, an assurance approach may support re-evaluation at runtime when the uncertainty related to inaccuracy and incompleteness is solved (at least partially).

**Competing criteria.** Assurance approaches should take into account possible competing criteria. While it is desirable to optimize the utility of evidence (coverage, quality) provided by an assurance approach, it is important that the approach allows making a tradeoff between the utility of the evidence it provides and the costs for it (time, resource consumption).

**User interaction.** Assurance is related also to the way users interact with the system, in particular with respect to changing request load and changing user profiles/preferences. For instance, when the request load for a service increases, the as-sured level of availability of that service may be reduced. Self-adaptation may reconfigure the internal server structure (i.e., adding new components) to maintain the required level of assurance. In the case of changing user preference/profile variations, the level of assurances may either have to be recomputed or they may be interpreted differently. An assurance technique should be able to support such variations in the ways the system is used.

**Handling alternatives** is another important benchmark criterion for assurance of self-adaptive systems. An assurance technique should be able to deal with a flat space of reconfiguration strategies but also with one obtained by composition. In this context an important capability is to support assurances in the case one adaptation strategy is pre-empted by another one.

## 6.2 Basis of Assurance Benchmarking

The second aspect of benchmarking techniques for perpetual assurances defines the basis of assurance, i.e., the reasons why the researchers believe that the technique makes valid decisions. We consider techniques based on historical data only, projections in the future only, and combined approaches. In addition, we consider human-based evidence as a basis for assurance.

**Historical data only.** At one extreme, an adaptation decision can be solely based on the past historical data, e.g., [12][17][28][29][33][64]. An issue with such adaptive systems is that there may be no guarantees that the analyses based on historical data will be reified and manifested in the future executions of system. Nevertheless, in settings where the past is a good indicator of the system's future behavior, such approaches could be applicable.

**Projections into the future only.** At the other extreme, a system may provide assurances that project into the future, meaning that the analysis is based on predictive models of what may occur in the future. The predictions may take different forms, including statistical methods e.g., ARIMA-based forecasting approaches [2][3][21] and fuzzy methods to address the uncertainty with which different situations may occur [30][38][51]. A challenge with this approach is the difficulty of constructing predictive models.

**Combined approaches.** Finally, in many adaptive systems, combinations of the aforementioned techniques are used, i.e., some past historical data is used together with what-if analysis of the implications of adaptations over the predicted behaviors of the system in its future operation.

**Human-based evidence.** Since users may be involved in the assurance process, we consider human-based evidence as a further basis for assurance. Input from users may be required before performing adaptations, which may be critical in terms of provided services. Consider a situation in which a self-adaptive service-oriented system needs to be reconfigured by replacing a faulty service. Before substituting this service with a new one, selected by an automated assurance technique, the user may want to check its SLA to confirm its suitability as a replacement according to a criterion that cannot be easily applied automatically. Such a criterion may be subjective, not lending itself to quantification. Thus, by means of accepting the SLA of the new service, the user provides a human-based evidence to complement the assurances provisioned through an automated technique, which can be based on past historical data, predictive models or a combination of both.

**Discussion.** When publishing research results, it is important to explicitly discuss and benchmark the basis of assurances. For instance, if historical data is used for making decisions, the research needs to assess whether past behavior is a good indicator of future behavior, and if so, quantify such expectations. Similarly, if predictive models are used for making decisions, the researchers need to assess the accuracy of the predictions, ideally in a quantitative manner.

### 6.3 Stringency of Perpetual Assurances

A third important benchmarking aspect of techniques for perpetual assurances of self-adaptive systems is the nature of rationale and evidence of the assurances, which we term "stringency of assurances." The rationale for assurances may differ greatly depending on the purpose of the adaptive system and its users. While for one system the only proper rationale is a formal proof, for another system, proper rationale may be simply statistical data, such as the probability with which the adaptation has achieved its objective in similar prior situations. For example, a safety-critical system may require a more stringent assurance rationale than a typical consumer software system. Note that the assurance rationale is a criterion that applies for different phases of adaptation (monitoring, analysis, planning, and execution). Different phases of the same adaptive system may require different level of assurance rationale. For instance, while a formal proof is a plausible rationale for the adaptation decisions made during the planning phase, it may not be appropriate for the monitoring phase. Similarly, confidence intervals for the distribution of monitored system parameters may be a plausible rationale for the monitoring phase, but not necessarily for the execution phase of a self-adaptive software system.

### 6.4 Performance of Approaches to Provide Perpetual Assurances

Finally, for an assurance approach to be applicable in real-world scenarios it is essential to evaluate its performance, which is the fourth and final proposed benchmark aspect. We consider a set of benchmarks, which directly follow from efficiency and scalability concerns discussed in Section 4, namely timeliness, computational overhead, and complexity analysis.

**Timeliness.** The time required to achieve the required evidence is a key performance benchmark criteria, which is obvious for techniques that are used at runtime.

**Computational overhead.** Related to timeliness is the computational overhead, i.e., the consumed resources (e.g., memory and CPU) for enacting the assurance approach.

**Complexity analysis.** As each assurance technique has an (implicit) complexity that may affect its applicability in terms of the analysis required for a problem at hand, we propose a complexity analysis benchmark. Assurance techniques may be benchmarked in terms of their scope of applicability across different types of problems.

The following table summarizes the benchmark aspects, and for each of them the different benchmark criteria.

**Table 4 Summary benchmark aspects and criteria**

| Benchmark Aspect | Benchmark Criteria | |
|---|---|---|
| | Criteria | Description |
| Capabilities of approaches to provide assurances | Variability | Capability of an approach to handle variations in requirements (adding, updating, deleting goals), and the system (adding, updating, deleting elements) |
| | Inaccuracy & incompleteness | Capability of an approach to handle inaccuracy and incompleteness of models of the system and context |
| | Competing criteria | Capability of an approach to balance the tradeoffs between utility (e.g., coverage, quality) and cost (e.g., time, resources) |
| | User interaction | Capability of an approach to handle changes in user behavior (preferences, profile) |
| | Handling alternatives | Capability of an approach to handle changes in adaptation strategies (e.g., pre-emption) |
| Basis of assurance benchmarking | Historical data only | Capability of an approach to provide evidence over time based on historical data |
| | Projections in the future | Capability of an approach to provide evidence based on predictive models |
| | Combined approaches | Capability of an approach to provide evidence based on combining historical data with predictive models |
| | Human evidence | Capability of an approach to complement automatically gathered evidence by evidence provided by humans |
| Stringency of assurances | Assurance rational | Capability of the approach to provide the required rational of evidence for the purpose of the system and its users (e.g., completeness, precision) |
| Performance of approaches | Timeliness | The time an approach requires to achieve the required evidence. |
| | Computational overhead | The resources required by an approach (e.g., memory and CPU) for enacting the assurance approach. |
| | Complexity | The scope of applicability of an approach to different types of problems. |

Some of the benchmark criteria from **Table 4** directly reflect a specific requirement from **Table 2**. For example, the 'Timeliness' criterion is directly linked to requirement R8 ('Produce timely updates'). Other criteria relate to multiple requirements from **Table 2**. As an example, the 'Human evidence' criterion links to R5 ('Integrate new evidence'), R7 ('Provide evidence for human activities that realize R5'), and R10 ('Auditable arguments'). Finally, some arguments only link indirectly to the requirements from **Table 2**. This is the case for the 'Handling alternatives' criterion, which relates to the solution for self-adaptation and this solution may provide different levels of support for the requirements of perpetual assurances.

# 7 Example Case

We now present a case that supports the evaluation, comparison, and ranking of approaches for perpetual assurances. The example case has been conceived to challenge the capability of approaches for self-adaptation to achieve perpetual provisioning of assurances for different requirements driven by the benchmark criteria discussed in the previous section. The particular ways in which different approaches solve the challenges proposed by the example case offers two distinct advantages. First it allows comparing prototypal research efforts, and second it acts as a testbed to demonstrate the effectiveness of the different techniques adopted by the different self-adaptive solutions.

We start with motivating the domain of the case and provide a set of general adaptation scenarios. The scenarios are linked to the types of uncertainty occurring in self-adaptive systems and the requirements for perpetual assurances. Next we describe the concrete case. We then give generic adaptation scenarios and explain benchmark criteria. Finally, we illustrate a comparison of two approaches for a concrete scenario.

### 7.1 Domain and General Adaptation Scenarios

The case comes from the domain of service-based systems, in which software services offered by third-party providers are dynamically composed at runtime to deliver complex functionality. Service-based systems are widely used in e-commerce, online banking, e-health and many other types of applications. They increasingly rely on self-adaptation to cope with the uncertainties that are often associated with third-party services [6][13][14][20][64] as the loose coupling of service-oriented architectures makes online reconfiguration feasible. Hence, the example case is a prototypical application.

A typical service-based system consists of a composition of web services that are accessed remotely through a software application called a workflow engine. Several providers may offer services that provide the same functionality, often with different levels of performance, reliability, costs, etc. Service providers can register concrete services at the service registry. The workflow of a composite service finds addresses (end points) of concrete services via the service registry. The way in which a workflow engine employs concrete services in order to provide the functionality required by the user is specified in the workflow that the engine is executing.

**Table 5** lists a set of generic adaptation scenarios for service-based systems. These scenarios provide increasing challenges to self-adaptation in general and the provision of perpetual assurances in particular. They are not meant to be exhaustive, but cover typical types of adaptation problems with different types of uncertainties.

Adaptation scenarios differ based on the characteristics summarized below.

**Types of uncertainties that the system needs to handle:** different types of uncertainty present different challenges to adaptation solutions and to the approaches used for perpetual assurances; for example monitoring or quantifying uncertainty (R1 and R2 in Table 2) associated with the failures of an individual service may be less challenging than integrating a new type of service that becomes available.

**Types of requirements that the self-adaptive system must meet:** self-adaptation can be used to achieve different types of system requirements.[2] Scenarios with a single requirement are typically less challenging for assurance approaches than scenarios where multiple competing requirements need to be balanced. Different combinations of requirements pose different challenges to perpetual assurances; e.g., monitoring, quantifying, and integrating new evidence (R1, R2, and R5) may be more demanding for requirements that require real-time tracking, while managing overlapping uncertainty resources (R3) and combining new evidence (R6) may be particularly challenging for collecting evidence of interdependent requirements.

Table 5 Generic adaptation scenarios for service-based systems

| Scenario | Type(s) of uncertainty | Type(s) of requirements | Observable properties | Types of adaptations (examples) |
| --- | --- | --- | --- | --- |
| S1 | Individual service failure | Reliability | Success or failure of each service invocation (Boolean – true=success, false=failure) | Select equivalent service, invoke idempotent services in parallel, enact alternative service |
| S2 | Variation of quality-of-service property over time (e.g., response time) | Quality of service (e.g. performance) | Variations in the quality of service property (e.g. changes in the response time for each service invocation (ms)) | Select equivalent service, invoke idempotent services in parallel, enact alternative service |
| S3 | New alternative service becomes available | Reliability, performance, cost | Available services for each operation (Set of currently available services) | Select new concrete service, enact new service |
| S4 | New type of service becomes available, requirement for new functionality | Add new service | Request to add new functionality (String), available concrete services including services for the new functionality (Set of currently available services) | Adapt workflow, enact adapted workflow, select new concrete service, enact new service |

**Observable attributes/properties:** the observations of the system and its execution context (R1 to R7) that are required depend on scenario, and introduce different challenges for adaptation solutions and approaches for perpetual assurances. For example monitoring the failures of a service is less demanding than observing the emergence of a new type of service (which may not have been anticipated upfront) and handling the request for integrating it into the system.

**Types of adaptations that are possible:** different scenarios require different types of adaptations. As an example, handling service failures may involve trying equivalent services until an invocation completes successfully, while introducing a new type of service requires an adaptation of the service workflow. On the other hand, approaches may be available that offer different adaptation strategies; e.g., an alternative strategy for reducing the likelihood of an operation failing may be executing multiple idempotent services in parallel (and using the result returned by the successful execution that completes first).

---

[2] It is important to distinguish the requirements that the self-adaptive system needs to realize (reliability, performance, etc.) and the requirements for approaches of perpetual assurances.

Different approaches to providing perpetual assurance should be evaluated and compared based on:

1. The benchmark aspects and criteria from **Table 4**.
2. Their ability to handle the scenarios in **Table 5** (or a subset of them).
3. The quality of the adaptation for which they provide assurance, which can be evaluated as described in the existing literature, e.g., [60].

### 7.2 Tele Assistance System

We now present the case. The concrete application is a Tele Assistance System (TAS) that offers health support to patients using home devices. The example case was presented as an exemplar at SEAMS 2015; for details we refer to [63]. A concrete realisation of the application is available at the community website.[3] TAS is based on the example introduced in [6] and later used in [13][29][56]. TAS offers a composite service that uses the following services:

- Alarm Service, which provides the operation sendAlarm
- Medical Analysis Service, which provides the operation analyzeData
- Drug Service, which provides the operations changeDoses and changeDrug

Multiple providers offer concrete services for the TAS with different characteristics, e.g. for reliability and cost. Each TAS implementation may use a particular strategy to select concrete services, e.g., based on the minimum response time.

TAS executes the workflow shown in **Figure 1**. The system uses the sensors embedded in a wearable medical device to takes periodical measurements of the vital parameters of a patient, and invokes the Medical Analysis Service for their analysis. The analysis result may trigger the invocation of the Drug Service to deliver new medication to the patient or to change his/her dose of medication, or the invocation of the Alarm Service, leading to an ambulance being dispatched to the patient. The Alarm Service can also be invoked directly by the patient by means of a panic button on the wearable device.

### 7.3 Adaptation Scenarios and Benchmark Criteria

We now provide instances of the generic adaptation scenarios from Table 5.

**Setting:** The TAS can use one of several concrete services for each abstract service it requires. To select concrete services the workflow uses a service registry with service descriptions that is refreshed from time to time. A service description lists the maximum failure rate and maximum response time promised by the provider of the concrete service, and the cost per service invocation.

---
[3] https://www.hpi.uni-potsdam.de/giese/public/selfadapt/exemplars/tas/

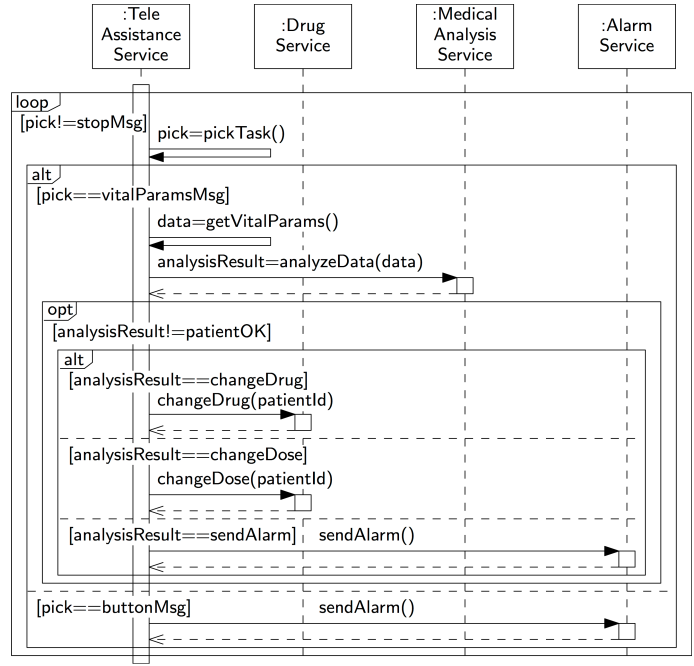

**Figure 1. Workflow of the TAS [63]**

**S1 - Individual service failure**

**Without adaptation**: Concrete service i that provides the sendAlarm operation is selected such that:

(i) $fr_i = \min_{1 \leq j \leq n}(fr_j)$, where $fr_j$ is the promised maximum failure rate of the j-th concrete alarm services,

(ii) $cost_i$ is the minimum cost of the concrete Alarm Service options that satisfy (i).

**Uncertainty:** The concrete services that provide the Alarm Service fail from time to time (we assume that there are at least two concrete alarm services).

**Adaptation requirement**: The failure rate of workflow executions that consist of an invocation of the Alarm Service does not exceed a predefined value.

**With adaptation:** The system observes each successful and failed Alarm Service invocation and dynamically selects the lowest-cost concrete service that meets the requirement.

**S2 - Variation of failure rate of services over time**

**Without adaptation**: A pair of concrete services that provide the Medical Analysis Service and Alarm Service is selected such that:

(i) $fr_1 + fr_2 \leq X$, where $fr_1$ and $fr_2$ are the promised maximum failure rates of the two selected services, and X is a pre-specified maximum failure rate of workflow executions comprising an invocation of the Medical Analysis Service followed by an invocation of the Alarm Service.

(ii) $cost_1 + cost_2$ is minimal cost across all combinations of concrete Medical Analysis Service and Alarm Service options that satisfy (i).

**Uncertainty:** The failure rates of the concrete services that provide the Alarm Service change significantly over time.

**Adaptation requirement**: The system requirement of a maximum failure rate of workflow executions comprising an invocation of the Medical Analysis Service followed by an invocation of the Alarm Service lower than or equal to X with minimal cost of service selections is maintained, regardless of the changing failure rates of concrete services over time.

**With adaptation:** The system observes each successful and failed invocation of a Medical Analysis Service and Alarm Service, uses these observations to estimate the actual failure rates of the concrete services, and dynamically selects the lowest-cost pair of concrete services that meets the requirement.

**S3** - New service becomes available

**Without adaptation**: A concrete service that provides the Alarm Service is selected as in Scenario S1. The registry with service descriptions is periodically refreshed with a period T.

**Uncertainty:** A new concrete service that provides the Alarm Service becomes available at some time t.

**Adaptation requirement**: The failure rate of workflow executions that consist of an invocation of the Alarm Service does not exceed a predefined value X.

**With adaptation:** The system observes each successful and failed invocation of an Alarm Service and dynamically selects the lowest-cost concrete service from the set of concrete alarm services, including, after refreshment of service descriptions, the new concrete service.

**S4** - New type of service becomes available

**Without adaptation**: A concrete service that provides the Alarm Service is selected as in Scenario S1.

**Uncertainty:** A new type of service called Inform Relatives becomes available at time t; the Inform Relatives service informs relatives of the patient in case of invocations of the Alarm Service; concrete services of the Inform Relatives service are characterized by a cost.

**Adaptation requirement**: After a time $\Delta t$ a concrete service that provides the Inform Relatives service is selected; the lowest cost concrete Inform Relatives service is invoked after every selection of an Alarm Service.

**With adaptation:** The system discovers new Inform Relatives services, it integrates the corresponding abstract service in the workflow, caches the concrete instances of the new service, and dynamically selects the lowest-cost concrete service from the set of Inform Relatives services.

For each of these scenarios, different benchmark criteria can be applied. For example, for Scenario S2, we may apply the benchmark variability (the capability of an ap-proach to handle variations in requirements) by comparing the percentage of time solutions assure compliance of the requirement that the maximal failure rate of workflow executions is not violated when different distributions of changing failure rates of services are imposed. In Scenario S4, we may benchmark incompleteness by comparing the capability of an approach to provide partial assurances that a new service for contacting relatives is correctly integrated in the workflow. We may apply benchmark performance of approaches for different scenarios by comparing the cost of different approaches to assure compliance with the requirement at runtime, in terms of overhead in time, CPU, and memory.

### 7.4 Concrete Example Scenario

To conclude this section we illustrate the comparison of two approaches for perpetual assurances for a concrete scenario.

**Concrete setup.** We use a scenario with an equal number of service instances for each of the three service types used by TAS. An overview of the profiles as declared by third-party service providers for a setup with five instances is shown in Table 6.

**Table 6 Third party service profiles for TAS scenario**

| Service ID | Alarm Service | | Medical Analysis Service | | Drug Service | |
|---|---|---|---|---|---|---|
| | Failure rate ($h^{-1}$) | Cost (¢) | Failure rate ($h^{-1}$) | Cost (¢) | Failure rate ($h^{-1}$) | Cost (¢) |
| 1 | 0.11 | 4.0 | 0.12 | 4.0 | 0.01 | 5.0 |
| 2 | 0.04 | 12.0 | 0.07 | 14.0 | 0.03 | 3.0 |
| 3 | 0.18 | 2.0 | 0.18 | 2.0 | 0.05 | 2.0 |
| 4 | 0.08 | 3.0 | 0.10 | 6.0 | 0.07 | 1.0 |
| 5 | 0.14 | 5.0 | 0.15 | 3.0 | 0.02 | 4.0 |

Table 7 shows the distributions of requests in the TAS scenario.

**Table 7 Distribution of requests in the TAS scenario (as a fraction of all requests)**

| Distribution user requests | | Distribution requests after analysis | |
|---|---|---|---|
| Checking vital signs | Emergency request | Drug service | Alarm service |
| 0.75 | 0.25 | 0.66 | 0.34 |

The service profile parameters and the distribution of requests are subject to uncertainty. Concretely, we added uncertainty to the failure rates of services and the distribution of requests based on a normal distribution.

**Requirements.** The concrete requirements that should be fulfilled are:

R1. The failure rate of invocations of TAS per hour should be below 0.02.

R2. The average cost per invocation of TAS should be below 8.0 ¢.

Hence, the scenario is a concrete instance of abstract scenario S1 (individual service failure) combined with scenario S2 (variation of failure rates of services over time).

**Approaches.** We apply and compare two approaches for perpetual assurance: runtime quantitative verification (RQV) [13] and runtime statistical model checking (RSMC) [42]. RQV applies model-based evaluation of qualities using probabilistic verification techniques. Concretely, the approach models the TAS service system as a discrete-time Markov chain and uses the PRISM model checker [46] at runtime to assure that service configurations are selected that comply to the requirements. The Markov model is parameterized by the configurable parameters of the service-based system (i.e., service profile parameters and distributions of requests to TAS) that are updated at runtime. During analysis the configurations that satisfy the quality requirements for the system are identified. The planner uses the analysis results to create a plan for adapting the configuration as needed. RQV assures that selected configurations comply with the requirements. RSMC on the other hand applies model-based evaluation of qualities using statistical verification techniques. The approach models the TAS systems with stochastic timed automata and uses the Uppaal-SMC model checker [24] at runtime to select service configurations that comply to the requirements. Similar to the Markov model used with RQV, the stochastic timed automata model is parameterized with service profile parameters and changing distributions of requests to TAS that are updated at runtime. RSMC checks the required properties of the model based on a sample set of simulations. RSMC uses statistical techniques to decide whether the system satisfies the property with some degree of confidence. During analysis the configurations that satisfy the quality requirements with a defined degree of confidence are identified. The planner uses these configurations to create a plan for adapting the configuration as needed. Thus, in contrast to exhaustive approaches, such as RQV, RSMC does not provide 100% guarantees, but an estimation, which is bound to a confidence interval. On the other hand, by setting the verification parameters, SMCR allows to tradeoff between the accuracy and confidence of the guarantees it provides with the system resources it requires.

The left-hand side of Figure 2 shows the failure rates obtained when RSMC was used for the concrete setup shown in Table 6 and Table 7. The boxplots show averages of 20 runs over 5 hours (with approximately 1000 invocations per hour). We used four different settings of verification parameters $E$ and $A$ (RSMC Strategy). The value of $E$ defines the approximation interval, i.e. $[p - E, p + E]$ with $p$ the probability that invocations of TAS fail, and $1 - A$ is the confidence level for the result. The figure shows that the setting with $A = 0.05$ and $E = 0.05$ guarantees no violations of the failure rate with a confidence level of 95%. The results with more relaxed parameter settings provide a less accurate result with lower confidence levels, but the mean values for all considered strategies remain under the requirement constraint of 2%. The right-hand side of the figure shows that the time required to compute the verification results decreases when the accuracy and confidence of the results is lowered. The approach enables balancing the tradeoff between computation time and accuracy/confidence of runtime model checking.

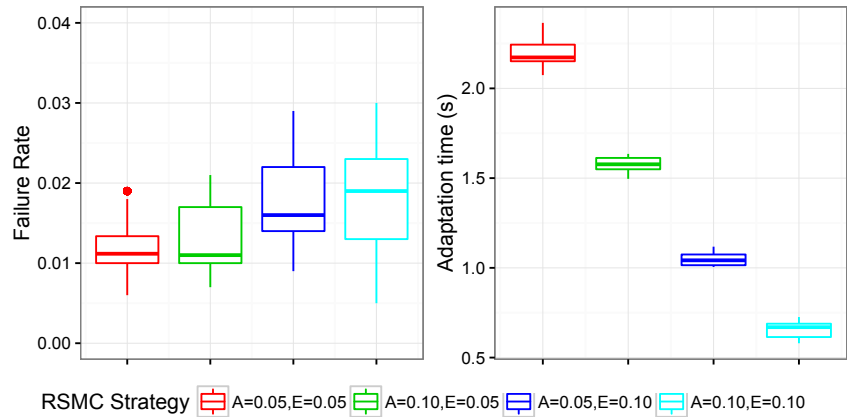

**Figure 2 Failure rates (left) and computation time (right) with RSMC**

We compared the adaptation time with RQV and RSMC for settings with an increasing number of service instances for TAS. Figure 3 shows the results of the experiments. The graphs confirm that RSMC is more efficient in required computation time and scales better. However, RQV can provide strict guarantees for the requirements at hand, while the efficiency of RSMC comes at an expense since fewer simulations can lead to results with less accuracy and confidence.

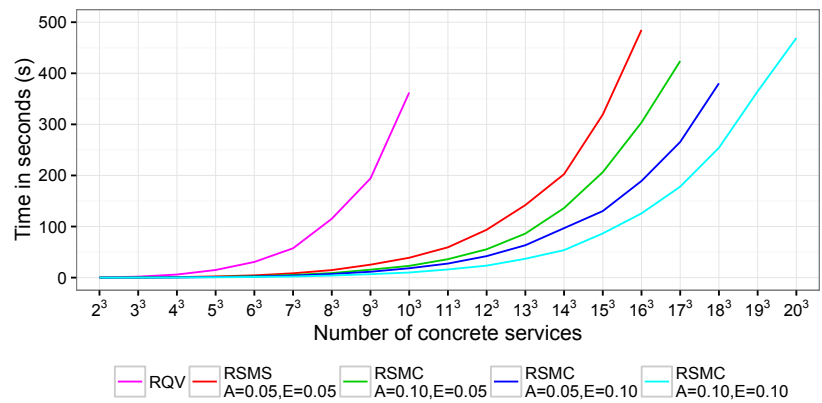

**Figure 3 Adaptation time for RQV versus RSMC**

We summarize the comparison for RQV and RSMC for a set of applicable benchmark criteria defined in **Table 4**.

**Table 8** Comparison benchmark criteria RQV versus RSMC

| Benchmark Criteria | | | |
|---|---|---|---|
| Criteria | Description | RQV | RSMC |
| Inaccuracy & Incompleteness | Capability of an approach to handle inaccuracy and incompleteness of models of the system and context | Yes through the use of a parameterized Markov model | Yes by using a parameterized stochastic timed automata with configurable verification parameters |
| Conflicting criteria | Capability of an approach to balance the tradeoffs between utility (correctness) and cost (verification time, memory) | Exact approach; no support to balance correctness versus verification time | Tradeoff between accuracy and confidence and required verification time and memory |
| Combined approaches | Capability of an approach to provide evidence based on combining historical data with predictive models | Both approaches provide evidence based on the prediction of expected behavior based on past observed data | |
| Timeliness | The time an approach requires to achieve the required evidence. | Exhaustive approach; expected to be limited to smaller modes than RSMC | Configurable based on setting of verification parameters |

## 8 Conclusions

Assuring requirements compliance of self-adaptive systems that have to operate under uncertainty calls for an enduring process where evidence is collected over the whole lifetime of the system. We coined this process as perpetual assurances for self-adaptive systems. We summarize the key challenges to realize perpetual assurances.

First, we need a better understanding of the nature of uncertainty for software systems and how this poses requirements for providing perpetual assurances. This paper provides an initial list of requirements for perpetual assurances based on a proposed classification of uncertainties. Additional research is required to test the validity and coverage of this set of requirements.

Second, we need a deeper understanding of how we can handle uncertainty in self-adaptive systems, and in particular, how can we monitor and quantify uncertainty. Currently, there is a growing understanding of how to handle uncertainty regarding parameters of the system, its goals, and the environment. However, how to handle uncertainty regarding parts of the system and the environment that may not be completely know upfront, or handling uncertainty regarding new goals remains to a large extent an open problem.

Third, deriving, integrating and combining new evidence pose additional hard challenges. A variety of techniques for obtaining evidence for requirements compliance of software systems exist. However, most of these techniques have been conceived for offline use. Perpetual assurance requires continuously deriving, integrating and combining new evidence, while the system is operating. We have presented decomposi-tion and model-based mechanisms than can potentially pave the paths to go forward. However, making these mechanisms effective is particularly challenging and requires a radical revision of many of the existing techniques.

Last but not least, to drive research on assurances for self-adaptive systems forward, we need good exemplars. Exemplars enable comparison of different solution, pinpoint the critical challenges, and demonstrate the effectiveness of the different mechanisms adopted by the self-adaptive solutions. The case used in this paper that is further explained in [63] provides one exemplar in the domain of service-based systems.